\documentclass[%reprint,
preprint,
]{revtex4-2}
\usepackage{graphicx}% Include figure files
\usepackage{dcolumn}% Align table columns on decimal point
\usepackage{bm}% bold math
\usepackage{xcolor}
\definecolor{red}{RGB}{200,0,0}
\newcommand{\lsb}{${\rm LaSb_2}$}
\usepackage{amsmath} % or simply amstext

\begin{document}

\title{Band Structure and Fermi Surface Nesting in LaSb$_2$}

\author{Evan O'Leary$^{1,2}$, Lin-Lin Wang$^{1,2}$, Yevhen Kushnirenko$^{1,2}$, Ben Schrunk$^{1,2}$, Andrew Eaton$^{1,2}$, Paula Herrera-Siklody$^{1}$, Paul C. Canfield$^{1,2}$ }
 \altaffiliation[ ]{canfield@ameslab.gov}
\author{Adam Kaminski$^{1,2}$}%
 \email{adamkam@ameslab.gov}
\affiliation{$^{1}$Iowa State University, Department of Physics and Astronomy, Ames, IA, 50014}

\affiliation{$^{2}$Ames Laboratory US Department of Energy, Ames, Iowa 50011, USA}

\date{\today}

\begin{abstract}
We use high-resolution angle resolved photoemission spectroscopy (ARPES) and density functional theory (DFT) to investigate the electronic structure of the charge density wave (CDW) system \lsb. This compound is among an interesting group of materials that manifests both a CDW transition and lower temperature superconductivity. We find the DFT calculations to be in good agreement with our ARPES data. The Fermi surface of \lsb\ consists of two small hole pockets close to $\Gamma$ and four larger pockets near the Brillouin zone (BZ) boundary. The overall features of the Fermi surface do not vary with temperature. A saddle point is present at $-0.19$ $eV$ below the Fermi level at $\Gamma$. Critical points in band structure have more pronounced effects on a materials properties when they are located closer to the Fermi level, making doped \lsb\ compounds a potential interesting subject of future research. Multiple peaks are present in the generalized, electronic susceptibility calculations indicating the presence of possible nesting vectors. We were not able to detect any signatures of the CDW transition at 355 K, pointing to the subtle nature of this transition. This is unusual, given that such a high transition temperature is expected to be associated with the presence of a large CDW gap. This is confirmed through investigation of the Fermi surface and through analysis of momentum distribution curves (MDC). It is possible that changes are subtle and occur below current sensitivity of our measurements.

\clearpage

\end{abstract}

\maketitle

%\tableofcontents

\section{Introduction}
\noindent
The rare earth diantiminides have been known crystallographically since 1969, but until the systematic growth of single crystals from excess Sb-solution little was known about their physical properties \cite{eatough1969high,canfield1992growth,canfield1991novel}. The RSb2 (R = La - Sm) series was first studied in single crystal form in with subsequent papers studying effects of doping, pressure and high magnetic field \cite{pcanfieldanisotropicmagnetic,bud2023rapid}. The crystal structure of \lsb\ is orthorhombic and highly layered, with a minor difference between the in-plane lattice constants \cite{galvis2013scanning}. \lsb\ can be described as metallic-micaceous in that it readily exfoliates into highly malleable 2-D like sheets\cite{pcanfieldanisotropicmagnetic}. As such it is very well suited for ARPES experiments. 
\lsb\ is part of a special group of materials that exhibit superconductivity and charge density wave (CDW) order \cite{galvis2013scanning,ruszala2020dirac,luccas2015charge}. The coexistence of superconductivity and CDW order have been studied previously in the transition metal dichalcogenide systems ${\rm NbSe_2}$, ${\rm 1T-TaS_2}$, and ${\rm TaSe_2}$ \cite{lian2018unveiling,sipos2008mott,lian2019coexistence}. Similarly to \lsb\, they are also highly layered. The superconducting transition in \lsb\ occurs at T$_c$ = 1.2 K, while two separate CDW transitions are proposed at 355 K and 13 K \cite{galvis2013scanning,ruszala2020dirac,luccas2015charge}. Another interesting feature of \lsb\ is the large, linear magneto resistance that does not saturate at high fields \cite{galvis2013scanning}.

%Talk about superconductivity and transport properties here.(2) 
The transport properties of \lsb\ have been extensively studied. The resistivity of \lsb\ is expected to be anisotropic, based on the crystal structure. This was confirmed by single crystal four-probe resistivity measurements \cite{pcanfieldanisotropicmagnetic}. A more recent study claims that the single crystal \lsb\ has an isotropic resistivity tensor \cite{fischer2019transport}. The unexpected isotropic resistivity tensor is explained by crystal imperfections. Further, scanning tunneling microscopy measurements show that \lsb\ enters a superconducting state at T$_c$ $\approx$ 1.2 K, while more recent resistivity measurements place the transition at T$_c$ $\approx$ 0.4 K \cite{galvis2013scanning,bud2023rapid}. Unfortunately, this temperature is so far unattainable for ARPES measurements.  

%find sources for CDW intro(1) https://doi.org/10.1016/0370-1573(85)90073-0 talk about how disorder pins incommesurate CDW                         
Several experiments have attempted to verify the CDW state in \lsb\ below 355 K, with evidence remaining fairly elusive \cite{galvis2013scanning, luccas2015charge, fischer2019transport, bud2023rapid}. Resistivity data indicates CDW order in \lsb\ through a dip and a hysteresis of ~2 K at around 350 K  \cite{luccas2015charge,bud2023rapid}. The CDW transition is more apparent and shifts to roughly 150 K upon substitution of La with Ce to around 50\% \cite{luccas2015charge}. The sharpening of the hysteresis is suspected to be caused by the introduced disorder in the system. Single crystal X-ray diffraction on pure \lsb\ showed peaks where systematic absences were expected, indicating a mixing of the a- and b- directions potentially as a result of twinning or stacking faults \cite{fischer2019transport}. This is potential evidence that innate disorder in pure \lsb\ may be the origin of the CDW state. The resistivity hysteresis may also be enhanced by applied pressure, increasing to a width of ~50 K, and shifting to lower temperature until it is completely suppressed at ~6-7 kbar \cite{bud2023rapid,weinberger2023pressure}. 

%Check wording in the paper (this is what is stated in the paper) ,talk about magnetoresistance
Another interesting property of \lsb\ is the large, linear magnetoresistance that does not saturate up to a field of 45 T \cite{galvis2013scanning,weinberger2023pressure, luccas2015charge}. It is speculated that scattering rate variations with applied magnetic field are the cause of the linear magnetoresistance \cite{ditusa2011optical}. This is based on the minor changes seen in the optical reflectivity when the magnetic field is varied. The large linear magnetoresistance makes \lsb\ a good candidate for magnetic field sensors \cite{young2003high}.   

%talk about previous ARPES data, topological nodal structure
The electronic band structure of \lsb\ has been previously investigated using de Haas-van Alphen (dHvA) technique, ARPES , and density functional theory calculations \cite{goodrich2004haas,acatrinei2003angle}. The dHvA measurements showed a quasi-two-dimensional Fermi surface with two nearly cylindrical extremal orbits, and one small ellipsoidal orbit \cite{goodrich2004haas}. These measurements were in rough agreement with the calculated Fermi surface. The DFT calculations seemed to agree well with the observed Fermi surface measured above the proposed CDW transition. The latest ARPES measurements showed that \lsb\ has multiple topological nodal points, lines and surfaces, including eightfold degenerate nodal points \cite{qiao2022multiple}. The multiple topological nodal structure is cited as another potential explanation of the large linear magnetoresistance in this material. We are not aware of any previous measurements of the band structure near the CDW transition at 355 K, motivating this study.

\section{Methods}
\noindent
%Crystal growth method description
Large single crystals of \lsb\ were grown out of antimony flux \cite{pcanfieldanisotropicmagnetic,canfield1992growth,bud2023rapid}. Elemental La
+(99.99+ Ames Laboratory) and Sb (99.999+ Alfa Aesar) were combined in the ratio of
${\rm La_5Sb_{95}}$ and placed into the bottom part of a 2 ml Canfield Crucible set (CCS) \cite{canfield2016use,crucible}. The CCS was sealed into an amorphous silica tube with silica wool on top of the CCS to act as a cushion during the decanting process. The sealed ampule was heated over 3 hours to 1100°C, held at 1100°C for 5 hours and then cooled to 1000°C over 1 hour. After sitting at 1000°C for 5 hours, the ampoule was cooled to 675°C over 99 hours. Upon reaching 675°C, the ampule was removed from the furnace and decanted in a lab centrifuge \cite{canfield2019new}. After cooling to room temperature, the ampule was opened, revealing large, sometimes crucible limited crystals of \lsb.

%ARPES description here
The ARPES measurements were performed by cleaving \lsb\ \textit{in situ} at base pressure lower than 5 $\times$ 10$^{-11}$ Torr. The 330 K and 360 K data were measured from the same sample which was cleaved at 330 K. The 7 K data was measured and cleaved at 7 K. A Scienta DA30 analyzer was used, in conjunction with a tunable, picosecond Ti:Sapphire laser and fourth harmonic generator operating at 6.7 $eV$ \cite{jiang2014tunable}. A second, fixed wavelength laser was also used in conjunction with the fourth harmonic generator and operated at 7 $eV$. The chemical potential was determined by measuring a polycrystalline Cu reference sample that was in electrical contact with the \lsb, and fitting a Fermi function to the data. One data set, shown in Fig.~\ref{fig:DFT} (c) was taken while applying a -7 V bias potential to the sample. A copper disk of radius 1.2 cm was electrically isolated from the rest of the ARPES system and the potential was supplied by a DC Power Supply 9116 from BK Precision directly to the disk. This technique has been tested and theoretically analyzed, where for a comparable setup there were little to no distortions when the applied bias is greater than -10 V \cite{gauthier2021expanding}.

%DTF calculation section 
Band structures of \lsb\ have been calculated in density functional theory \cite{hohenberg1964inhomogeneous_DFT1,kohn1965self_DFT2} (DFT) using PBE \cite{perdew1996generalized_DFT3} exchange-correlation functional with spin-orbit coupling (SOC) and experimental lattice parameters. DFT calculations were performed in VASP \cite{kresse1996efficient_DFT5,kresse1996efficiency_DFT6} with a plane-wave basis set and projector augmented wave \cite{blochl1994projector_DFT4} method. We used the primitive unit cell of 12 atoms with a $\Gamma$-centered Monkhorst-Pack \cite{monkhorst1976special_DFT8} (11x11x3) $k$-point mesh and a Gaussian smearing of $0.05$ $eV$ to converge charge density. A tight-binding model based on maximally localized Wannier functions \cite{marzari1997maximally_DFT9,souza2001maximally_DFT10,marzari2012maximally_DFT11} was constructed to reproduce closely the bulk band structure including SOC in the range of $E_F \pm 1 eV$ with La $sdf$ and Sb $p$ orbitals. Then the spectral functions and Fermi surface of a semi-infinite \lsb\ (001) surface were calculated with the surface Green’s function methods \cite{lee1981simple_DFT12,lee1981simple_DFT13,sancho1984quick_DFT14,sancho1985highly_DFT15} as implemented in WannierTools \cite{wu2018wanniertools_DFT16}. The real part of susceptibility function has been calculated on a dense k-point mesh of (160x160x40) to study Fermi surface nesting.

\section{Results and discussion}
\begin{figure}
\centering
\includegraphics[scale=0.5]{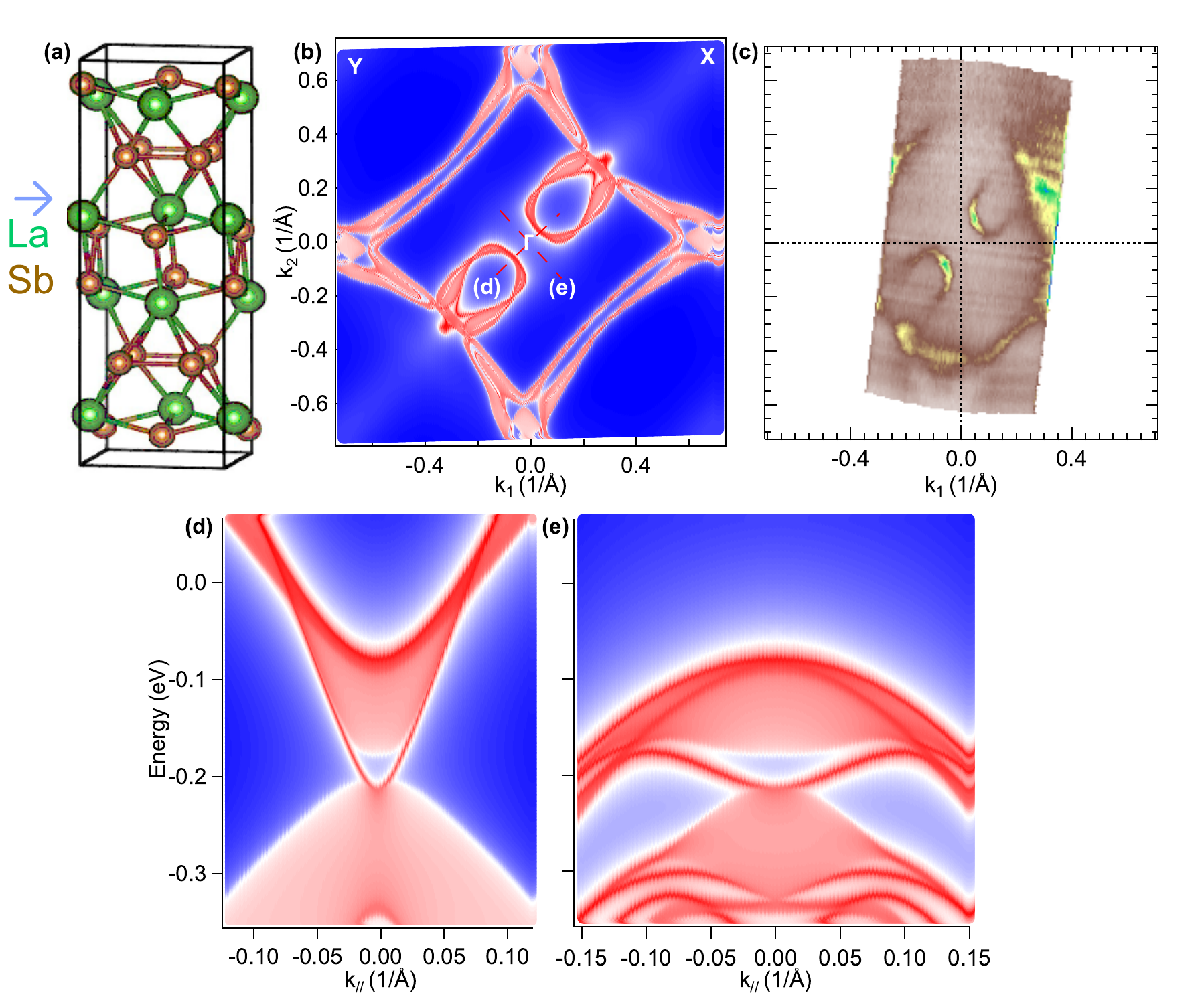}
    \caption{Density functional theory band structure calculation, with predicted surface termination for \lsb. (a) Crystal structure of \lsb, arrow indicates termination plane. (b) Predicted Fermi surface for (LaSb)$_2$ termination, showing BZ labeling. (c) ARPES intensity at Fermi level taken at 7 K integrated within 10 $meV$ of the Fermi cutoff. ARPES data was taken with a bias potential of -7 V applied to the sample. (d),(e) Energy dispersion cuts, compare with Fig.~\ref{fig:SaddlePoint} (e),(h) respectively.}
    \label{fig:DFT}
\end{figure}

\begin{figure}[!htb]
\centering
\includegraphics[scale=0.47]{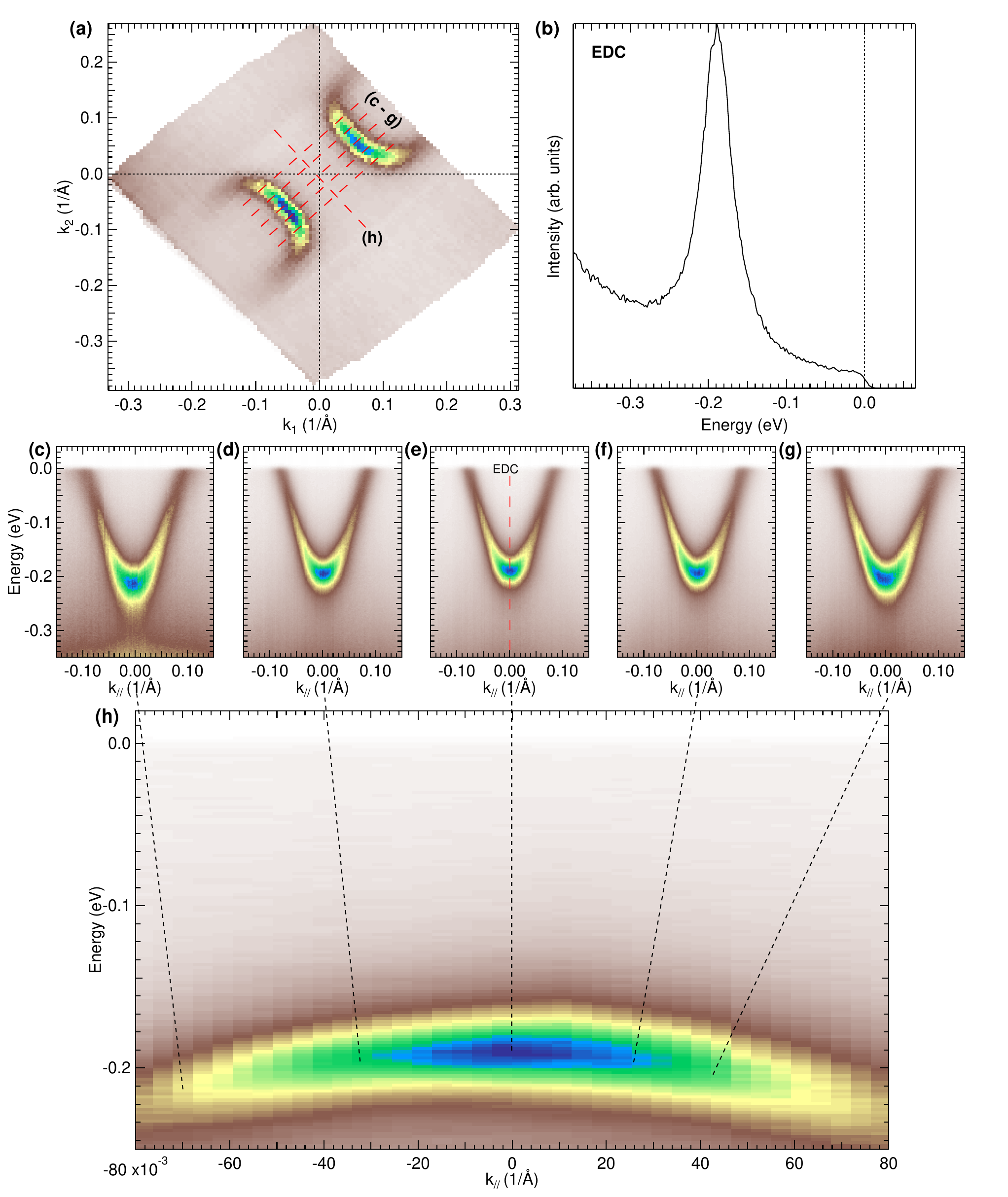}
    \caption{ARPES data showing saddle point. (a) Map shown at 7 K integrated within 10 $meV$ around the Fermi cutoff. (b) EDC taken at cut shown in panel (e) with peak indicating the location of the saddle point. (c)-(h) Energy dispersion cuts through lines in panel (a) showing a saddle point in the band structure. }
\label{fig:SaddlePoint}
\end{figure}

\noindent
Density functional theory calculations are presented in Fig.~\ref{fig:DFT}. The crystal structure and the proposed cleaving plane, indicated by the arrow, is shown in Fig.~\ref{fig:DFT} (a). The termination occurs at the {${\rm (LaSb)_2}$} layer. Comparing Fig.~\ref{fig:DFT} (d),(e) with  Fig.~\ref{fig:SaddlePoint} (e),(h) respectively, shows this termination gives a calculated band structure that has good agreement with our ARPES data. Both show a saddle like dispersion centered on $\Gamma$. Comparison of Fig.~\ref{fig:DFT} (b) and (c), reveals that two small round pockets appear in both the DFT calculations and ARPES data and have very similar shape. There are multiple outer pockets predicted by the calculations, which are also seen in the ARPES data. They are similar in shape and location, but ARPES intensity fades out in the upper corner of Fig.~\ref{fig:DFT} (c). The ARPES data appear to have good agreement with the DFT calculations. 
%Paragraph 2: Talk about critical point in the band structure, including why they are interesting, how they occur, what they mean

Critical points in the energy dispersion of 3-dimensional materials produces a divergence in the density of states (DOS), and causes the derivative of the DOS to be non-continuous, known as a van Hove singularity. When critical points in electronic structure are near or coincide with the Fermi energy, materials may present with interesting properties such as superconductivity, magnetism, or charge and spin density waves \cite{mcchesney2010extended,li2010observation}. It is general practice to try and tune the location of the van Hove singularity to a lower binding energy. This can be accomplished either via doping or other mechanical methods as in twisted graphene  \cite{li2010observation,mcchesney2010extended}. In Fig.~\ref{fig:SaddlePoint} we present ARPES data that reveals the presence of a saddle point located at $\Gamma$. The portion of the pockets discussed previously that are closest to $\Gamma$ make up the upwards dispersion of the saddle point. The dispersions across the pockets for several cuts are shown in Fig.~\ref{fig:SaddlePoint} (c)-(g), showing an upwards dispersion.  Fig.~\ref{fig:SaddlePoint} (h) shows a cut perpendicular through $\Gamma$ and shows a downwards dispersion. In Fig.~\ref{fig:SaddlePoint} (c)-(g), the bottom of the band is exactly the dispersion shown in Fig.~\ref{fig:SaddlePoint} (h). The energy distribution curve (EDC) shown in Fig.~\ref{fig:SaddlePoint} (b), indicates the location of the saddle point is $-0.19$ $eV$. Given the saddle point occurs at a rather high binding energy, it is unlikely to be tuned to the Fermi level. It is worth recalling that the CDW was enhanced as \lsb\ was doped with Ce \cite{luccas2015charge}. This enhancement could potentially be an effect of the saddle point shifting lower in binding energy.

%Paragraph 3: Talk about data above and below CDW

Charge density wave states occur when electrons and phonons become coupled, causing the electron density to be periodically displaced. The period of the CDW can be either commensurate or incommensurate with the period of the lattice. A CDW that is commensurate with the lattice can cause small displacements of the ions that compose the lattice. This disrupts the periodicity of the lattice, leading to changes in the size and shape of the BZ and potentially causing band folding. A CDW that is incommensurate with the lattice is generally much weaker in its affects. Incommensurate CDWs may interact with impurities in the lattice, leading to a pinning of the phase of the CDW. Strong pinning can cause more pronounced affects such as lattice distortions and large frequency dependent responses in the conductivity \cite{gruner1985charge}. A CDW transition often leads to subtle or less subtle changes in the electronic structure, such as band back folding caused by new periodicity, and or development of new energy gaps due to Fermi surface nesting like what is seen in the CDW material ${\rm 2H-NbSe_2}$ \cite{borisenko2009two,rahn2012gaps}. We looked carefully for these effects in the portion of the BZ accessible in our experiments. 

\begin{figure}[!htb]
\centering
\includegraphics[scale=0.83]{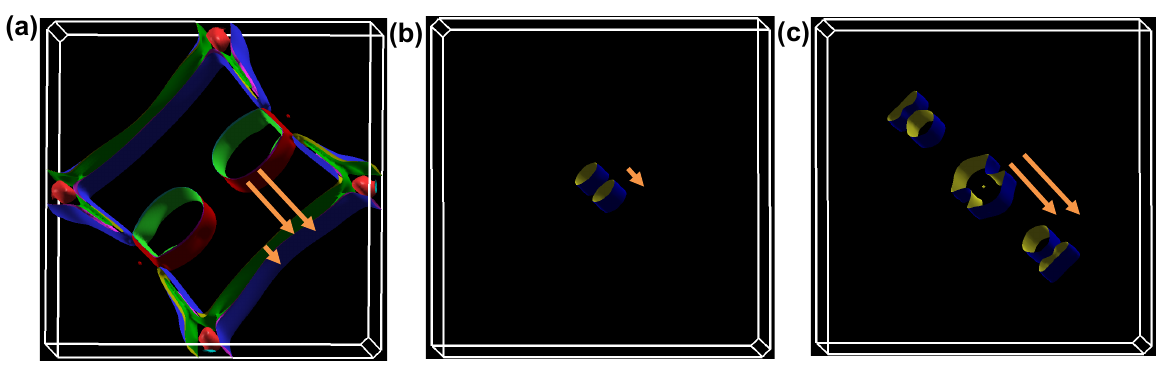}
    \caption{Density functional theory electronic susceptibility calculation. (a) Calculated band structure at the Fermi level. (b),(c) Electronic susceptibility shown in momentum space, (b) shows largest peak in electronic susceptibility, (c) shows second largest peak in electronic susceptibility. Predicted nesting vectors are shown by arrows.}
\label{fig:ElectronicSusceptibility}
\end{figure}

Proposed nesting vectors predicted by DFT calculations are shown in Fig.~\ref{fig:ElectronicSusceptibility} (a). The lengths of the nesting vectors were determined through investigation of the electronic susceptibility, shown in Fig.~\ref{fig:ElectronicSusceptibility} (b) and (c). A peak in the electronic susceptibility may help to satisfy the Chan-Heine criterion for strong electron-phonon coupling, increasing the probability of CDW formation \cite{xu2021topical}. The nesting vectors connect parallel contours on the calculated Fermi surface. The calculations provide insights on where potential effects due to the CDW might be observed in ARPES data.

\begin{figure}    
\centering
\includegraphics[scale=0.53]{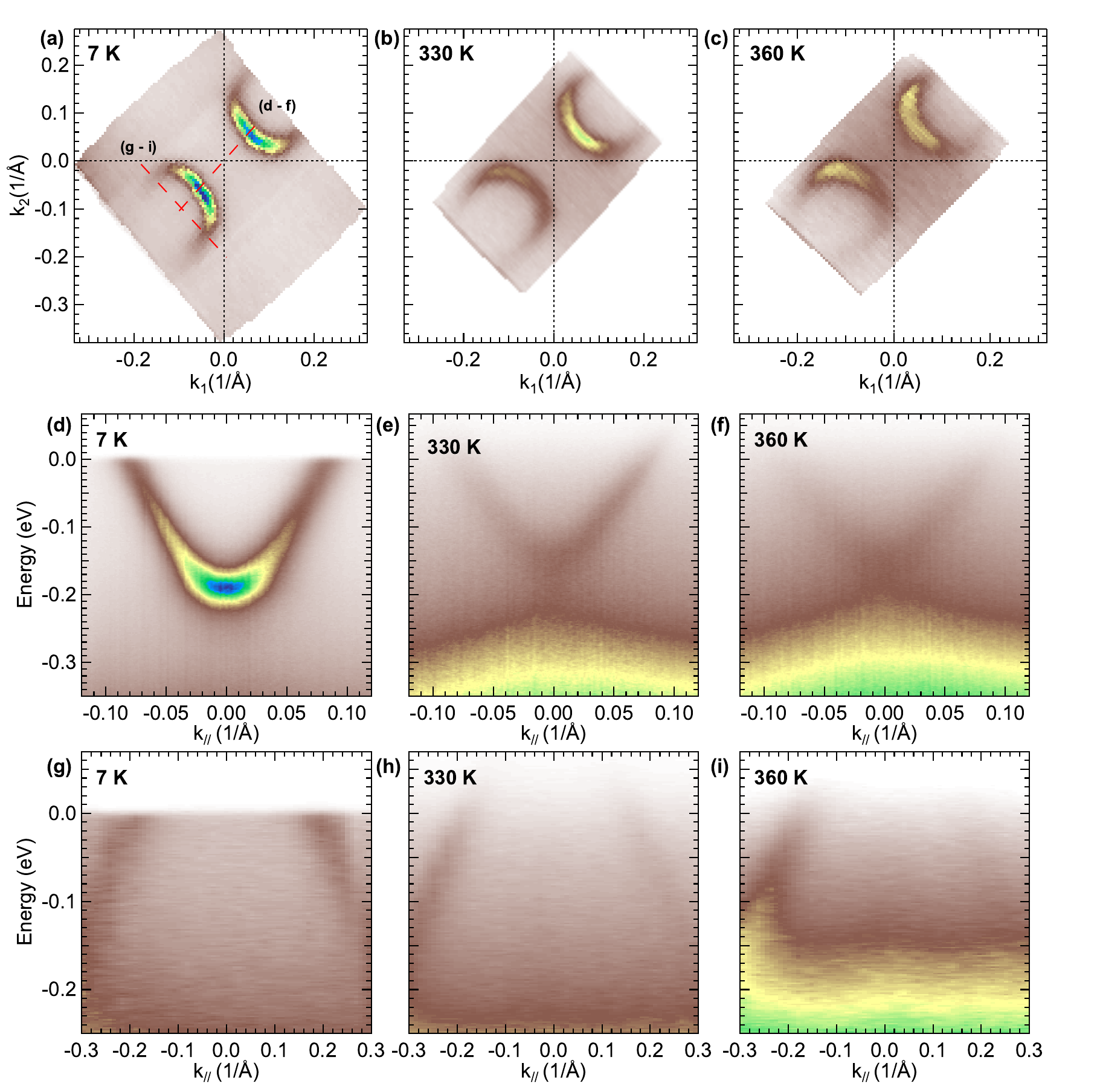}
    \caption{ARPES intensity at the Fermi level, at several temperatures across the CDW transition. (a)-(c) Measured at 7 K, 330 K, and 360 K respectively. (d)-(i) Band dispersion at cuts shown by red lines in panel (a).}
    \label{fig:TempComp}
\end{figure}

\begin{figure}[!htb]
\centering
\includegraphics[scale=0.6]{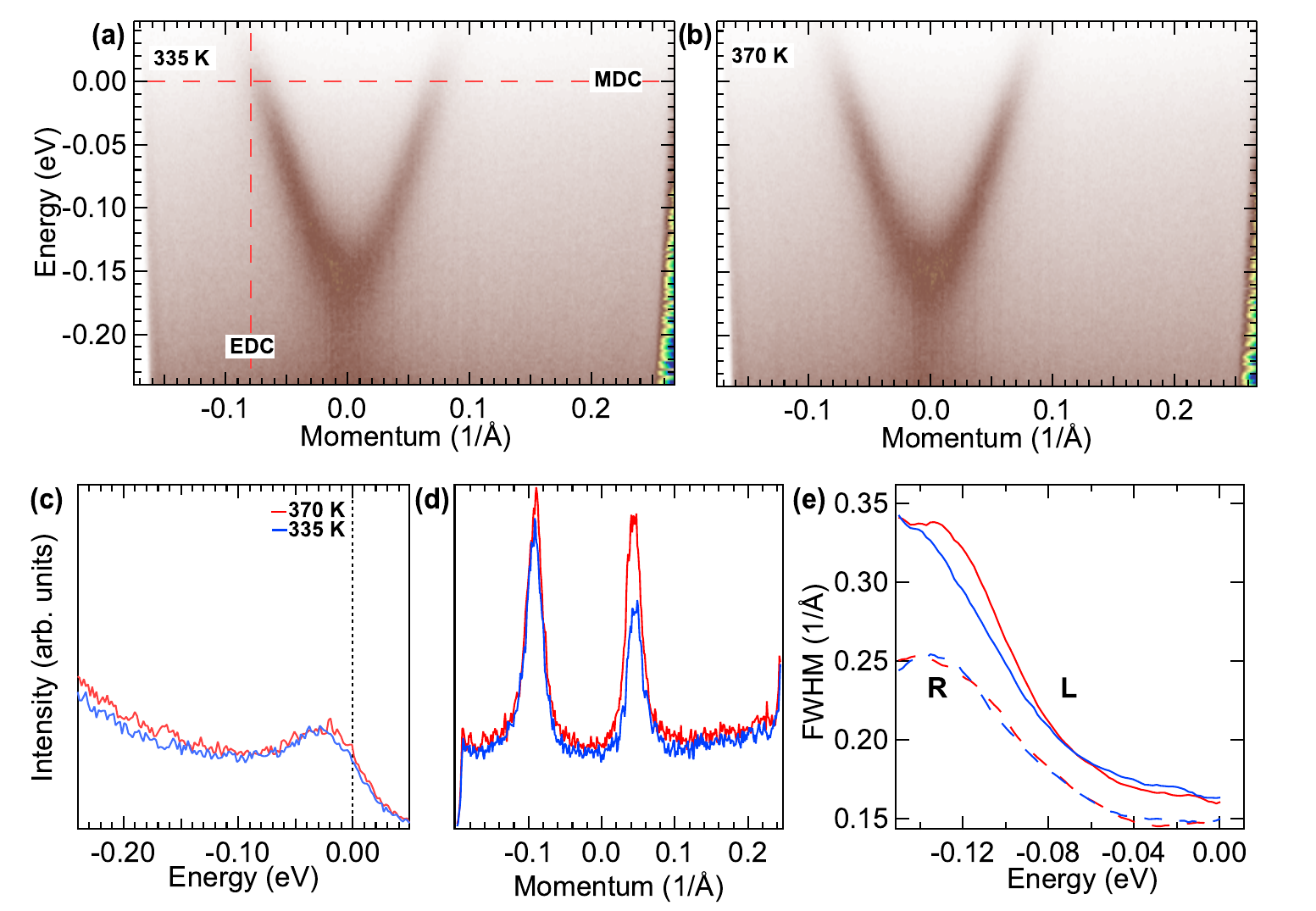}
    \caption{Comparison of bandwidth when crossing through the CDW transition. (a),(b) Cuts taken of electron band centered at $\Gamma$, at 335 K and 370 K. (c),(d) Energy distribution curve and momentum distribution curve, respectively, through lines in panel (a). (e) Full width at half maximum (FWHM) of the Lorentzian fit of MDCs, solid line is FWHM of the left branch of the band, the dashed line is the right branch. Color indicates temperature.}
\label{fig:WidthComp}
\end{figure}

In Fig.~\ref{fig:TempComp} (a)-(c), the temperature evolution of the Fermi Surface (FS) is shown. The only noticeable difference in the ARPES data through the transition is a slight shift in intensity of the spectra. This can be seen in the cuts along the X-$\Gamma$-X direction shown in Fig.~\ref{fig:TempComp} (d),(e). The band appears sharper and more intense at 7 K, with high intensity at  $-0.19$ $eV$. The spectrum at 330 K shows a band with the same shape to the band in the 7 K spectra, but with a much lower intensity. At 330 K the band has slightly shifted to lower binding energy, relative to its position at 7 K. The shift of the band is an effect of changing temperature and not caused by the CDW transition. In Fig.~\ref{fig:TempComp} (h),(i) we examine one of the locations that DFT calculations predict possible nesting vectors may form. From these data, we see no changes indicative of Fermi surface nesting. An EDC taken at the band crossing is shown in Fig.~\ref{fig:WidthComp} (c), there is no indication of an energy gap forming when entering the CDW state. Given the high transition temperature, one would expect an energy gap to be easily detectable. We show MDCs at the Fermi level in Fig.~\ref{fig:WidthComp} (d), again no observable changes occur when going through the CDW transition. To more precisely investigate any potential changes due to the CDW, we studied changes to the quasi-particle lifetime across the CDW transition. This was done by fitting the MDC, like the ones shown in Fig.~\ref{fig:WidthComp} (d), with Lorentzian functions for a range of binding energies. As shown in Fig.~\ref{fig:WidthComp} (e), there are no quantitative changes in the width of the electron bands when going through the CDW transition, demonstrating that the CDW did not cause any changes in the quasi-particle lifetime in this portion of the BZ.

\section{Conclusions}

In summary, we preformed detailed ARPES measurements and DFT calculations to investigate the electronic properties of \lsb\ that hosts high and low temperature CDW states. The ARPES data agrees relatively well with DFT calculations and shows a Fermi surface with two small pockets located near $\Gamma$, and four larger pockets around the edge of the BZ. These results agree with previous dHvA data, ARPES data and DFT calculations \cite{goodrich2004haas,acatrinei2003angle}. No visible changes in the band structure occur when going through the CDW transitions. The most interesting feature was a saddle point predicted by DFT calculations and confirmed by our ARPES measurements. The saddle point was centered at $\Gamma$ and $-0.19$ $eV$ below the Fermi level. This is relatively high in binding energy, but worth investigating further through doped \lsb\ compounds.

\clearpage

\clearpage

\bibliography{apssamp}% Produces the bibliography via BibTeX.

\end{document}